# Parametric Modeling of EEG Signals

**Rakesh K. Sharma, Pradip Sircar***

Department of Electrical Engineering
Indian Institute of Technology Kanpur
KANPUR 208016, India
*Corresponding Author; Email Address: sircar@iitk.ac.in

## Abstract

In this paper, a new signal model is suggested for parametric representation of the electroencephalogram (EEG) signals. The proposed model which is an amplitude and frequency modulated sinusoidal signal model, has been found to capture the non-stationary characteristics of the EEG signal with good accuracy. When the EEG signal is considered for longer duration of time, the model parameters have turned to be time-variant.

## 1. Introduction

The electroencephalogram (EEG) signals capture the electrical activity of the brain as monitored by scalp electrodes [1]. A change in the electrical potential of a nerve cell in the brain is produced by ionic movement across the cell membrane in the course of the activity of a cell. The EEG signal level is typically 10 to 200 $\mu$ volts and occupies a frequency range from near dc to approximately 30 Hz. The classification of the EEG signals is based upon their frequency ranges. There are four fundamental rhythms namely, Alpha, Beta, Theta, Delta, found in an adult human brain that can be recorded.

The characteristics of EEG signal changes rapidly with change in the state of the mind, and are different for different age groups and subjects. These changes are in amplitude and frequency of four types of rhythm [2]. The Alpha rhythm is most prominent and useful of all rhythms.

A novel amplitude and frequency modulated sinusoidal (AFMS) model is proposed in this paper for representation of the EEG signals. Simulation with the Alpha rhythm shows the suitability of the AFMS model in representing the signal.

## 2. The AFMS Model

Let the discrete time random signal $x[n]$ be given as follows:

$$x[n] = A\cos[\omega_c n + k_f \sin(\omega_f n) + \theta] + (sAk_a/2)\cos[\omega_c n + \omega_a n + k_f \sin(\omega_f n) + \theta + \theta_a]$$
$$+ (rAk_a/2)\cos[\omega_c n - \omega_a n + k_f \sin(\omega_f n) + \theta - \theta_a - \theta_b] \quad (1)$$

where $A$ is the amplitude of the carrier wave, $\omega_c$ is the carrier angular frequency, $\theta$ is the phase of the carrier assumed to be a random variable with distribution $U[0, 2\pi)$, $k_f$ is the index of frequency modulation and $\omega_f$ is the modulating angular frequency, $k_a$ is the index of amplitude modulation and $\omega_a$ is the modulating angular frequency, $\theta_a$ is the phase of the upper sideband, $\theta_b$ is the additional phase of the lower sideband, $s$ and $r$ are scaling factors for the upper and lower sidebands, respectively, and $n$ is the discrete time. Without loss of generality, it is assumed that the scaling factor, $r = -1$.

Computing the autocorrelation function (ACF) of the sequence, $r_x[n,l] = E\{x[n]x[n+l]\}$, where $E$ is the expectation operator, one finds that the ACF depends on both of time $n$ and lag $l$. This finding proves that the signal in (1) is a non-stationary signal. Moreover, if only one observation of the random sequence $x[n]$ is available, the ensemble average of the ACF can not be computed. These two features of the modeled signal $x[n]$ should be kept in mind when we derive the technique for its parameter estimation.

## 2.1 Parameter estimation

We introduce the product function (PF) $p_x[l]$ defined as

$$p_x[l] = \{x[n]x[n+l]\}_{n=-l/2} \quad \text{for} \quad l = 0, \pm 2, \pm 4, \cdots \quad (2)$$

and then with a change of variable, $l/2 = k$, we compute the sequence $\{p_x[k]; k = 0, \pm 1, \pm 2, \cdots\}$. It can be shown that the sequence $p_x[k]$ comprises of components with cluster of angular frequencies centered at $2\omega_c$, $(2\omega_c - \omega_a)$, $(2\omega_c + \omega_a)$, $(2\omega_c - 2\omega_a)$, $(2\omega_c + 2\omega_a)$, beside a dc component and components with angular frequencies $\omega_a$ and $2\omega_a$ [3]. Therefore, if we can identify the centre frequencies $(2\omega_c + 2\omega_a)$ and $(2\omega_c + \omega_a)$ of the highest frequency cluster and the second highest frequency cluster, respectively, then the carrier angular frequency $\omega_c$ and the amplitude modulating angular frequency $\omega_a$ can be computed. The center frequency of a cluster can be determined easily by observing the symmetry of the magnitude–spectrum of the sequence $p_x[k]$ around that frequency [3].

For determination of the modulating angular frequency $\omega_f$ of frequency modulation, we notice that the separation of two adjacent peaks in any cluster of frequencies in the spectrum of $p_x[k]$ is given by $\omega_f$. Therefore, by taking an average of number of such separations, an accurate estimate of $\omega_f$ can be determined.

Next, we rewrite (1) using the Bessel functions $J_m(k_f)$ and $J_m(-k_f)$ as

$$x[n] = \tfrac{A}{2}e^{j\theta}\sum_{m=-\infty}^{\infty}J_m(k_f)e^{j(\omega_c n+m\omega_f n)} + \tfrac{A}{2}e^{-j\theta}\sum_{m=-\infty}^{\infty}J_m(-k_f)e^{j(-\omega_c n+m\omega_f n)} \quad (3)$$

$$+ \tfrac{sAk_a}{4}e^{j(\theta+\theta_a)}\sum_{m=-\infty}^{\infty}J_m(k_f)e^{j(\omega_c n+\omega_a n+m\omega_f n)} + \tfrac{sAk_a}{4}e^{-j(\theta+\theta_a)}\sum_{m=-\infty}^{\infty}J_m(-k_f)e^{j(-\omega_c n-\omega_a n+m\omega_f n)}$$

$$+ \tfrac{rAk_a}{4}e^{j(\theta-\theta_a-\theta_b)}\sum_{m=-\infty}^{\infty}J_m(k_f)e^{j(\omega_c n-\omega_a n+m\omega_f n)} + \tfrac{rAk_a}{4}e^{-j(\theta-\theta_a-\theta_b)}\sum_{m=-\infty}^{\infty}J_m(-k_f)e^{j(-\omega_c n+\omega_a n+m\omega_f n)}$$

Notice that once the angular frequencies $\omega_c$, $\omega_a$, $\omega_f$ are known, (3) can be solved in the least-squares sense to compute the terms of the form $A_{ci} J_m(k_f)$ for $i = 1, \cdots, 6$, where $A_{c1} = \frac{A}{2} e^{j\theta}$, $A_{c2} = \frac{A}{2} e^{-j\theta}$, and so on.

In order to determine the frequency modulation index $k_f$, we compute the ratios of Bessel functions of different orders and same argument:
$\frac{J_0(k_f)}{J_1(k_f)}, \frac{J_0(k_f)}{J_{-1}(k_f)}, \frac{J_0(k_f)}{J_2(k_f)}, \frac{J_0(k_f)}{J_{-2}(k_f)}$, etc., and then by using an appropriate look up table, $k_f$ can be evaluated [4]. Once the value of $k_f$ is known, the complex amplitudes $A_{ci}$ for $i = 1, \cdots, 6$ can be computed, which in turn can be used to estimate the parameters $A, \theta, k_a, \theta_a, \theta_b, s$ of the modeled signal. Note that the scaling factor $r$ is set at $r = -1$ for convenience of estimation.

3. **Model Fitting of EEG Signal**

We demonstrate the suitability of the proposed model for representation of a natural EEG signal. The Cz lead configuration of EEG signal of a subject, 25 year old male, is used for this purpose. The data is sampled at 500 Hz, and the sample data is plotted in Fig. 1. The analog signal was bandpass filtered between 1 Hz and 20 Hz, and then after sampling digitally filtered between 0.2 Hz and 200 Hz (both second order Bessel filters). The discrete Fourier transform (DFT) plot of the sampled data is shown in Fig. 2.

We consider a block of 41 samples of the signal for estimation of parameters. The size of the block was decided by plotting the DFT for various data lengths and selecting the length which gave a regular DFT plot. Note that for the EEG signal to fit the AFMS model, its DFT must shown three clusters of peaks around angular frequencies $(\omega_c - \omega_a)$, $\omega_c$, $(\omega_c + \omega_a)$. However, we do not use DFT plot for estimating the angular frequencies of the modeled signal because these estimate will not be accurate.

Using a block of 41 samples of the EEG signal, we compute the product function $p_x[k]$ and extract all the frequencies contained in the sequence. For this purpose the sequence $p_x[k]$ is first fitted into a linear prediction (LP) model, and the prediction coefficients are evaluated using the modified covariance method [5]. The frequencies of the modeled signal are obtained by computing the zeroes of the prediction error filter (PEF) [5].

Once we estimate all the parameters of the modeled signal for a block of data, the data set can be regenerated using the estimated values of the parameters in the model. Figs. 3-5 show original and regenerated signals for three separate blocks of data. The comparison of the original and regenerated data segments clearly shows the suitability of the proposed model in representing the EEG signal in parametric form. However, it should be pointed out that the values of the parameters are changing from one data segment to the next, which proves that what we really need in this case is a time-variant model.

## 4. Conclusion

The approach presented here provides a means to represent the EEG signal in terms of few parameters of the proposed model. It will be worthwhile to study how these parameters change with the state of mind. The research can be extended to find out whether the ranges of the parameters can be related to the diagnosis of disorders in the activity of brain .Finally, it is to be noted that the parametric modeling of EEG signals will allow monitoring the activity of the brain together with its timing information with high accuracy and precision.

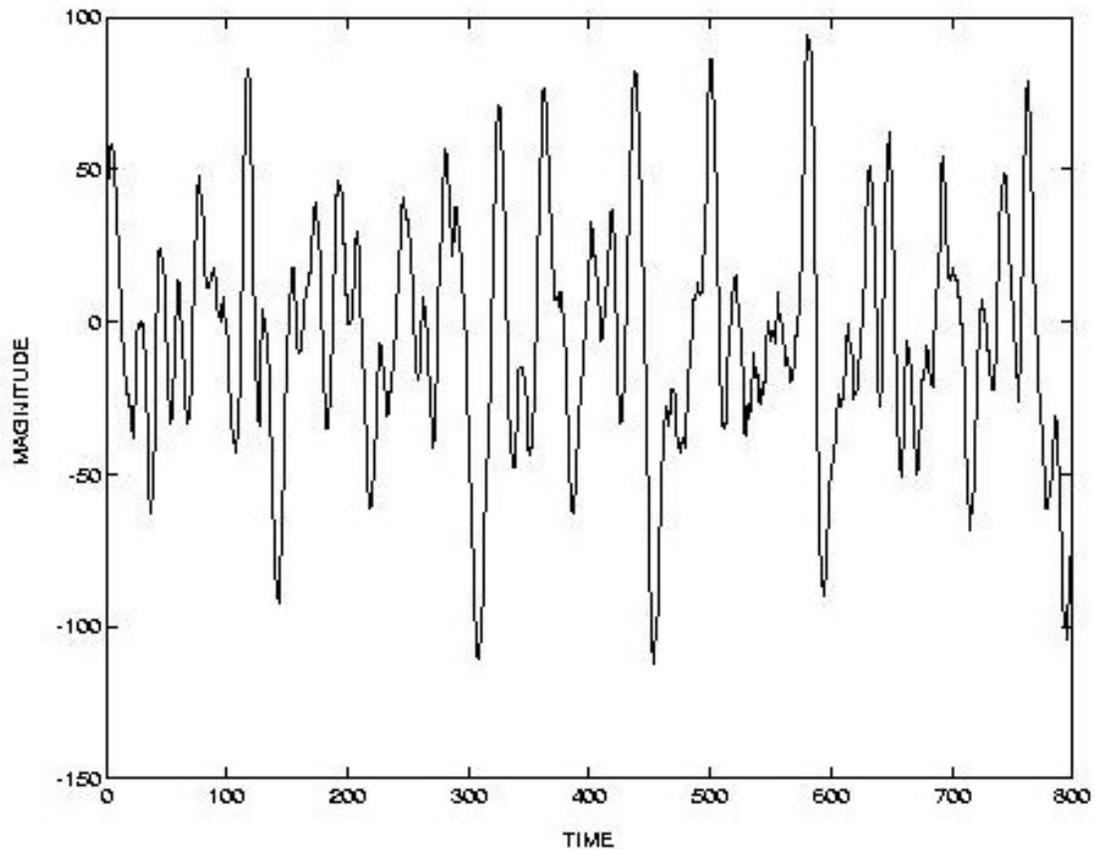

Fig.1 The EEG Signal

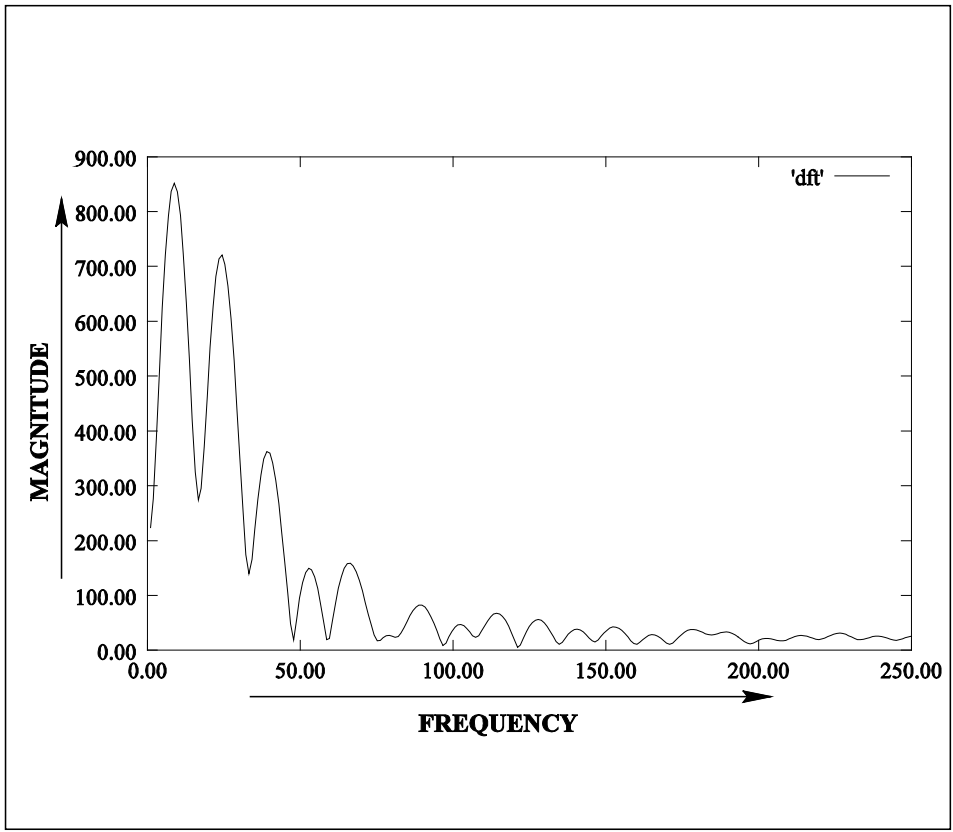

**Fig. 2 The DFT plot**

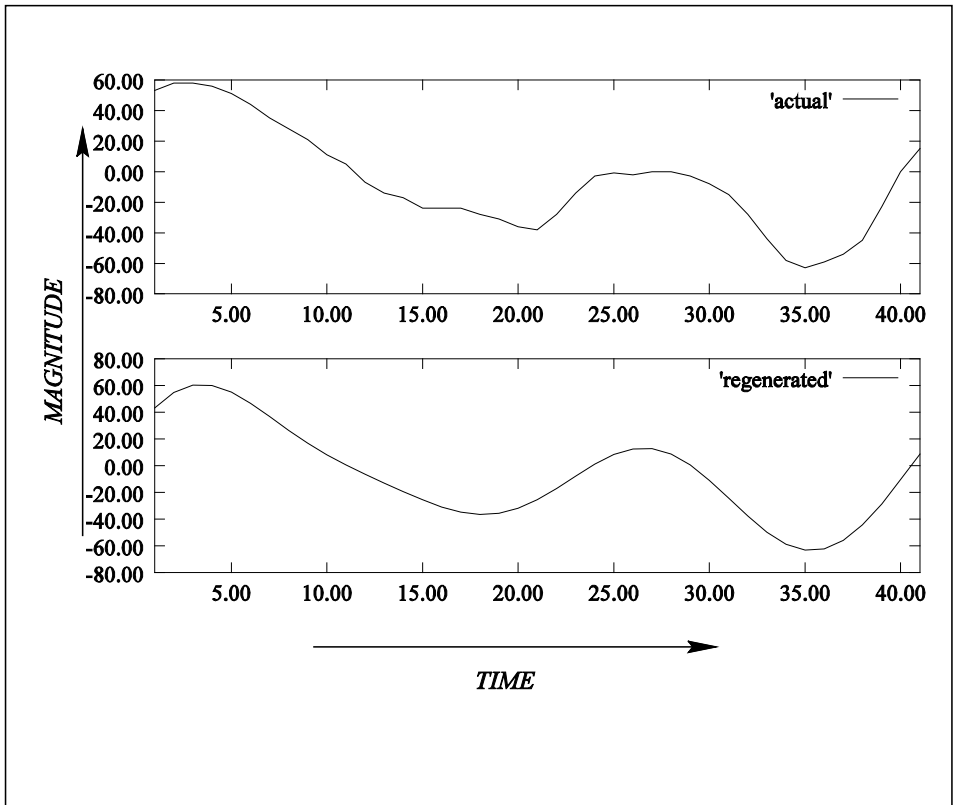

**Fig. 3 Regeneration of original signal 1**

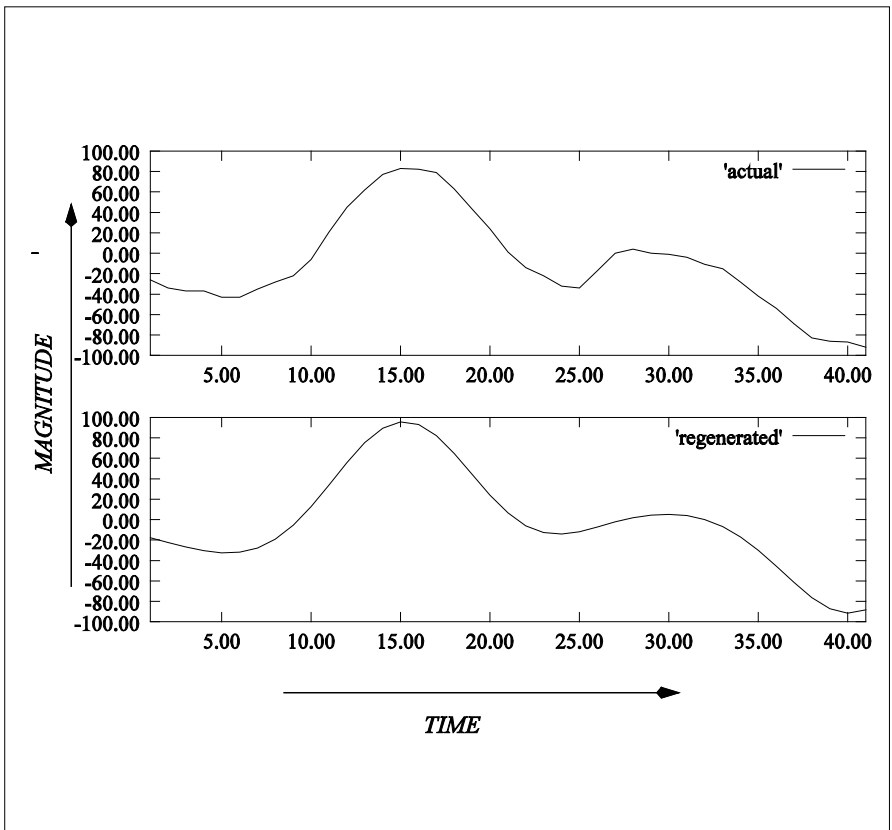

**Fig. 4 Regeneration of original signal 2**

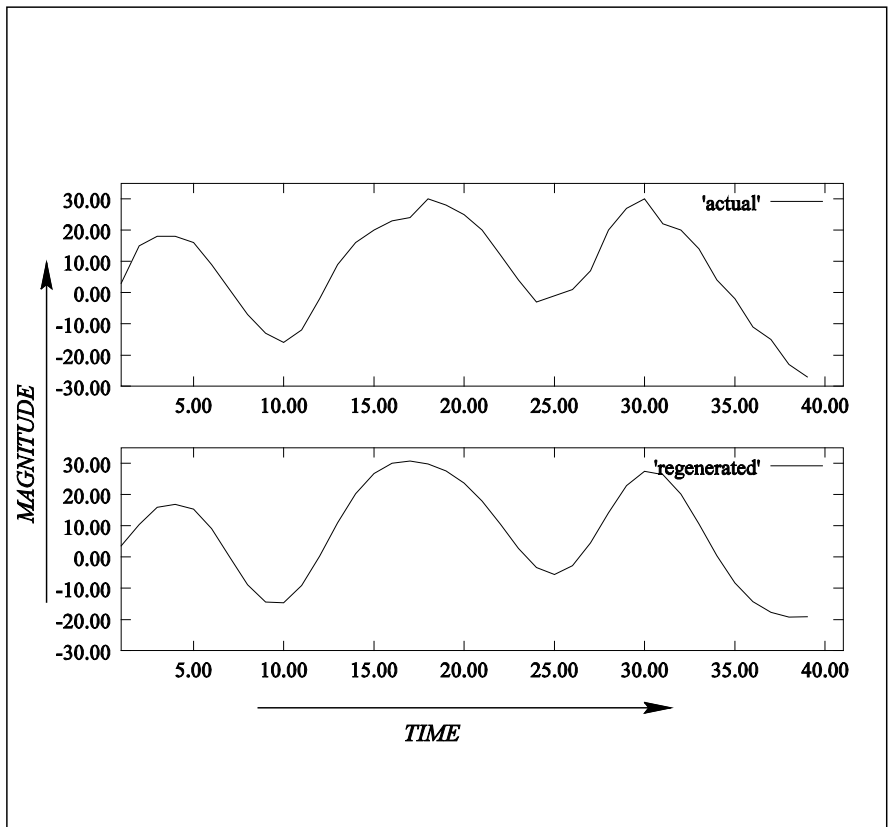

**Fig. 5 Regeneration of original signal 3**